\documentclass[aps,prl,twocolumn,showpacs,showkeys,superscriptaddress]{revtex4}

\usepackage{graphicx}

\begin{document}

\title{Current driven spin-wave instability triggered by the anomalous
Hall effect.}

\author{I.~Ya.~Korenblit}

\affiliation{The Raymond and Beverly Sackler School of Physics and
Astronomy, Tel Aviv University, Tel Aviv 69978, Israel}

\date{\today}

\begin{abstract}
We studied the effect of strong electric current on spin waves
interacting relativistically with the current. The spin-wave
spectrum is calculated at arbitrary direction of the wave vector.
It is shown that the alternating Hall current generated by the
alternating magnetic moment of the spin waves, reduces the
spin-wave damping. At strong enough unpolarized dc current  the
damping changes sign, and the spin-wave amplitude starts to
increase  exponentially fast  with time. The critical current for
the spin-wave instability is determined mainly by the anomalous
Hall effect, and can be much smaller than that for the spin-torque
mechanism of instability.
\end{abstract}

\pacs{75.30.Ds, 75.40.Gb, 75.47.-m}


\maketitle

 Current induced switching of the magnetization or spin wave excitation in
magnetically inhomogeneous systems, e.g. multilayers, has received
considerable attention during the last decade.\cite{TBB} When a
spin-polarized current passes through a ferromagnetic layer, the
electrons transfer  their spin angular momentum to the localized
spins of the ferromagnet resulting in a spin torque acting on the
magnetization.\cite{Sl1,Ber1} The theoretical predictions were
confirmed by several groups, the experiments being performed
mainly with trilayers of structure ferromagnet-normal
metal-ferromagnet, in which the layer magnetization may be
noncollinear, see Ref. \onlinecite{TBB} and references therein.

 It has been argued recently  that polarized current can
affect the magnetic properties also of an homogeneous bulk
ferromagnetic metal.\cite{BJZ,FBN,LZ,TSB} Adding the spin torque
linear in the {\it spin current}, to the Landau-Lifshitz equation
of motion, one gets a modified spin-wave (SW) spectrum, which
shows a current driven instability.
 In a
half-metal, when the density of the minority carriers is zero, the
spin current is equal to the electric current. In this case the
uniform ferromagnetic state becomes unstable at a critical current
given by the relation \cite{BJZ,FBN}
\begin{equation}
{\mathbf k}\cdot{\mathbf v_d}=\omega_{\mathbf k}, \label{kvd}
\end{equation}
where {\bf k} is the spin-wave wave vector, $\omega_{\mathbf k}$
is the spin-wave dispersion, and ${\mathbf v}_d$ is the electron
drift velocity, which is proportional to the electric current.
This "Doppler-shift" critical current is of order of $ 10^9$
A/cm$^2$,\cite{FBN} but in general, as has been shown by
Tserkovnyak {\it et al.},\cite{TSB} the critical current can be
strongly enhanced.

High enough current densities can excite SW excitations in
ferromagnetic layers even when the current is unpolarized, if the
source and drain contacts are nonsymmetric.\cite{PB} SW excitation
by an unpolarized current injected into a single ferromagnetic
film from a point contact was observed by Ji {\it et al.}
\cite{JCS} and considered theoretically in Refs. \onlinecite{XYO}.

 The original models by Slonczevski \cite{Sl1} and Berger \cite{Ber1}
  and all subsequent  considerations of
 current induced SW excitation rely on the exchange model of interaction
 between the itinerant electrons and the localized spins.
 In this paper we concentrate on the electromagnetic
  (relativistic) interaction of the electron current with the
field of the SW, which does not preserve the total spin.  We show
that in ferromagnetic conductors with large anomalous Hall effect
this interaction can lead to current induced SW instability at
critical {\it unpolarized} current of the same order or even
smaller than that in the exchange interaction models. Unlike the
exchange coupling of itinerant electrons with the SW, which is
effective only in the vicinity of the interface between normal and
ferromagnetic layers,\cite{Ber1} the above relativistic
interaction acts also in the bulk of the ferromagnet. Therefore,
the current induced SW instability cased by the Hall effect is not
restricted to a layered structure of the ferromagnet.

 The physical
mechanism for SW generation via the relativistic interaction is as
follows.
 Suppose for simplicity that an electric current {\bf j}$_0$ is driven parallel to
the magnetization {\bf M}$_0$. Consider a spin wave propagating
along {\bf M}$_0$. The oscillating magnetic moment and magnetic
field of the spin wave lie in the plane perpendicular to {\bf
j}$_0$ and {\bf M}$_0$. This gives rise to an alternating Hall
current perpendicular to {\bf M}$_0$, which in his turn creates a
magnetic field amplifying the field of the wave. If the electric
field is strong enough, the amplification will exceed the damping
due to eddy currents. If there are no other sources of damping,
the spin-wave system becomes unstable at such field. We show in
this paper that the instability is not restricted to the above
simple geometry. It takes place at any mutual orientation of {\bf
j}$_0$ and {\bf M}$_0$.

In ferromagnetic metals the main contribution to the Hall current
comes from the anomalous Hall effect caused by the spin-orbit
coupling in the metal. Since the anomalous Hall constant is by
orders of magnitude larger than the normal one, the critical
current for SW amplifying can be relatively small.

The full set of equations, which describes SW in a conducting
media interacting with an electric current consists of the
Landau-Lifsitz equation and of the Maxwell's equations. The
Landau-Lifshitz equation for the precession of the magnetic moment
is\cite{ABP}
\begin{equation}
{\partial{\mathbf m}\over \partial t}=\gamma{\mathbf
M_0}\times{\mathbf H}_{eff}+{1\over\tau_2}{\mathbf
H}_{eff}-{1\over\tau_1 M_0^2}{\mathbf M_0}\times({\mathbf
M_0}\times {\mathbf H}_{eff}). \label{LL}
\end{equation}
Here {\bf m} is the (small) transverse alternative part of the
magnetization, $m\ll M_0$, $\gamma$ is the gyromagnetic ratio,
$\tau_1$ and $\tau_2$ are phenomenological SW relaxation times,
and {\bf H}$_{eff}$ is the effective magnetic field given by
\begin{eqnarray}
{\mathbf H}_{eff}={\mathbf h}+\Bigl[{D\over\gamma M_0}\nabla^2
-{{\mathbf m_0}\cdot{\mathbf H_0}\over M_0}\nonumber\\
-K({\mathbf m_0}\cdot{\mathbf n})^2\Bigr]{\mathbf m} +K{\mathbf
n}({\mathbf n}\cdot {\mathbf m}), \label{He}
\end{eqnarray}
where {\bf h} is the alternative part of the magnetic field, $D$
is the stiffness constant of the SW, ${\mathbf H}_0$ is the
external magnetic field, which in restricted samples includes also
the demagnetizing fields,  {\bf n} and {\bf m}$_0$ are  unit
vectors directed along the anisotropy axis and the magnetization
respectively, and $K$ is the dimensionless anisotropy constant.

The Maxwell equations are:
\begin{eqnarray}
\nabla\times{\mathbf h} &=&{4\pi\over c}{\mathbf j},\nonumber\\
\nabla\times{\mathbf e} &=&-{1\over c}{\partial{\mathbf
b}\over\partial t}, \nonumber\\
\nabla\cdot{\mathbf b}&=&0. \label{max}
\end{eqnarray}
Here the alternating  magnetic induction is ${\mathbf b} ={\mathbf
h} +4\pi{\mathbf m}$, {\bf e} is the alternating electric field,
$c$ is the light velocity, and {\bf j} is the alternating electric
current given by \cite{GK}
\begin{equation}
{\mathbf j}=\sigma{\mathbf e} +\sigma(R_B{\mathbf
j_0}\times{\mathbf b} + R_M{\mathbf j_0}\times{\mathbf m}),
\label{J}
\end{equation}
where $R_B$ and $R_M$ are the ordinary and anomalous Hall
coefficients respectively, $\sigma$ is the conductivity, and
${\mathbf j}_0$ is the dc part of the electric current density. We
neglected in Eqs. (\ref{max}) the displacement current, since for
conductors considered here the inequality $\omega\ll\sigma$ always
holds. We skipped in Eq. (\ref{J}) a small term of order
$\sigma(R_B B_0 + R_M M_0)\ll1$. The SW frequences considered in
the paper are of order of $(10^9 - 10^{10}) s^{-1}$. i.e. much
smaller than all typical electron relaxation frequencies in a
ferromagnetic metal. Therefore, we used below the dc values of the
transport coefficients.

The Maxwell equations, with the current from Eq. (\ref{J}), relate
the Fourier transforms of {\bf h} and {\bf m} as
\begin{eqnarray}
{\mathbf h}=-4\pi[\omega_0k^2+i{2\over\delta^2}({\mathbf
k}\cdot{\mathbf v}_b-\omega)]^{-1}[\omega_0({\mathbf
k}\cdot{\mathbf m}){\mathbf k}\nonumber\\
-i{2\over\delta^2}({\mathbf k}\cdot{\mathbf m}){\mathbf v}_m
-i{2\over \delta^2}(\omega - {\mathbf k}\cdot{\mathbf v}){\mathbf
m}]. \label{hm}
\end{eqnarray}
Here $\omega_0=4\pi\gamma M_0$, the skin penetration depth,
$\delta$, at frequency $\omega_0$ is given by
$\delta=(c^2/2\pi\sigma\omega_0)^{1/2}$, and the effective
velocities ${\mathbf v}_b, {\mathbf v}_m$ and ${\mathbf v}$ are
related to the Hall coefficients by
\begin{equation}
{\mathbf v}_b=R_B c{\mathbf j}_0\equiv {\mathbf v}_d,~~~ {\mathbf
v}_m ={R_M c{\mathbf j}_0\over4\pi},~~~ {\mathbf v}={\mathbf v}_b
+{\mathbf v}_m.\label{vbm}
\end{equation}
It is supposed that $\delta$ is much larger than the electron mean
free path.

The Landau-Lifshitz equation yield another relation between {\bf
h} and {\bf m}. When ${\mathbf H}_0$ and ${\mathbf M}_0$ are
parallel and  directed along the anisotropy axis, this relation
reads:
\begin{eqnarray}
(\Omega_{\mathbf k}-\omega- i\alpha\Omega_{\mathbf k})m_{+}&=&
\gamma M_0h_{+}(1- i\alpha)\nonumber\\
(\Omega_{\mathbf k}+\omega+ i\alpha\Omega_{\mathbf k})m_{-}&=&
\gamma M_0h_{-}(1+ i\alpha).\label{mh}
\end{eqnarray}
Here
\begin{equation}
\alpha={1\over \gamma M_0}\left({1\over\tau_1} +
{1\over\tau_2}\right),\label{alt}
\end{equation}
 $m_{\pm}=m_x\pm im_y$, $h_{\pm}=h_x\pm ih_y$, while the axis $z$ is
along the magnetization. The frequency $\Omega_k$ is given by
\begin{equation}
\Omega_k=\gamma (H_0 +H_a) +Dk^2, \label{Om}
\end{equation}
where $H_a$ is the anisotropy field: $H_a= KM_0$.

Eqs. (\ref{hm}) and (\ref{mh}) give the dispersion relation for SW
in an external electric field. We consider in what follows
wave-vectors $k$, which are larger than $\delta^{-1}$:
$k\delta\gg1$. One obtains then to leading order in the small
parameter $1/k^2\delta^2$:
\begin{eqnarray}
\omega_{\mathbf k}^2(j_0)=\Omega_{\mathbf k}\Omega_{1{\mathbf
k}}+{2\omega_{\mathbf k}(0)\over\delta^2k^2}({\mathbf
k}\times{\mathbf v}_m)\cdot{\mathbf
m}_0\nonumber\\-i\Bigl[{2\over\delta^2k^2}(\Omega_{1{\mathbf
k}}+\Omega_{\mathbf k}\cos^2\theta)(\omega_{\mathbf k}(0)-
{\mathbf k}\cdot{\mathbf v})\nonumber\\ +\alpha\omega_{\mathbf
k}(0)(2\Omega_{\mathbf k} +\omega_0\sin^2\theta)\Bigr]\nonumber\\
-i{2\over\delta^2k^4} \Omega_{\mathbf k}{\mathbf k}\cdot{\mathbf
m}_0[({\mathbf v}_{m}\cdot{\mathbf k})({\mathbf m}_0\cdot{\mathbf
k})-k^2{\mathbf v_m}\cdot{\mathbf m}_0]. \label{om2}
\end{eqnarray}
Here $\theta$ is the angle between the wave vector {\bf k} and the
magnetization, $\Omega_{1{\mathbf k}}=\Omega_{\mathbf k}
+\omega_0\sin^2\theta$, and $\omega_{\mathbf k}(0) \equiv
\omega_{\mathbf k}(\sigma=0)=\sqrt{\Omega_{\mathbf
k}\Omega_{1{\mathbf k}}}$. In the absence of the current $j_0$,
this equation gives the usual  spectrum of SW decaying due to
Landau-Lifshitz-Gilbert damping $\alpha$, and  due to the  eddy
currents. The last decay is proportional to $1/k^2\delta^2$.

In what follows we consider such wave vectors that the
contribution to the SW damping from the last term in Eq.
(\ref{om2}) is equal to zero. This happens specifically, if {\bf
k} is along or perpendicular to the magnetization, or at any
$\theta$ provided {\bf k} is along the current. Eq. (\ref{om2})
then yields:
\begin{eqnarray}
\mbox{Re}\,\omega_{\mathbf k}(j_0)&=&\sqrt{\Omega_{\mathbf
k}\Omega_{1{\mathbf k}}}+{({\mathbf k}\times{\mathbf
v}_m)\cdot{\mathbf
m}_0\over\delta^2 k^2}, \nonumber\\
\mbox{Im}\,\omega_{\mathbf k}(j_0)&=&-(\nu_{\mathbf k}
+\alpha_{\mathbf k}) \left(\omega_{\mathbf k}(0) -{\nu_{\mathbf
k}\over\nu_{\mathbf k} +\alpha_{\mathbf k}}{\mathbf
k}\cdot{\mathbf v}\right). \label{okv}
\end{eqnarray}
Here
\begin{eqnarray}
\nu_{\mathbf k}&=& {\Omega_{1{\mathbf k}} +\Omega_{\mathbf
k}\cos^2\theta\over k^2\delta^2\sqrt{\Omega_{\mathbf
k}\Omega_{1{\mathbf k}}}},\nonumber\\
\alpha_{\mathbf k}&=&{\alpha\over 2\sqrt{\Omega_{\mathbf
k}\Omega_{1{\mathbf k}}}}(2\Omega_{\mathbf k}
+\omega_0\sin^2\theta). \label{nua}
\end{eqnarray}
 Eq. (\ref{okv}) is our main result. It shows that Im\,$\omega$
 changes sign at a critical velocity, $\mathbf{v}_c$, given by the relation
\begin{equation}
{\mathbf k}\cdot{\mathbf v}_c=\left(1+{\alpha_{\mathbf k}\over
\nu_{\mathbf k}}\right)\omega_{\mathbf k}(0). \label{vom}
\end{equation}
At higher effective velocities, i.e. at higher currents the
amplitude of spin waves, with {\bf k} satisfying Eq (\ref{vom}),
increases exponentially with time.

When $R_M/4\pi$ is smaller than $R_B$, i.e. $v=v_b=v_d$, the SW
instability condition (\ref{vom}) resembles those, obtained in
Refs. \onlinecite{BJZ,FBN,LZ,TSB} for SW instability in
half-metals triggered by spin-transfer torques, see  Eq.
(\ref{kvd}). However, $R_M$ in ferromagnetic conductors is usually
by many orders of
 magnitude
 larger than $R_B$. We have, therefore, $v\approx v_m\gg v_d$, and
 the critical current for the instability considered here is much
 smaller than the critical current given by Eq (\ref{kvd}).

 Measurements \cite{MAF} performed on Fe, Co, Ni and Gd films with the
 thickness of 1 $\mu$m, show that in pure metals, with the resistivity
 $\rho=(10^{-4}-10^{-6})~ \Omega$cm, the anomalous
 Hall conductivity $\sigma_H$ is of order $10^3 (\Omega$cm)$^{-1}$.
 Thus, for metals with the resistivity $\rho= (10^{-4} - 10^{-5})
 \Omega$cm, and with $M_0\approx 10^3$ G, one gets $R_M=
 (10^{-8} - 10^{-10})~ \Omega$cm/G. The values of $R_M$ for Ni films, which
  follow from the data obtained in Ref. \onlinecite{Gerber}, also  fall in
  this region.  It follows than from Eq (\ref{vbm}) that
the effective velocity $v_m$ is of order $v_m = (10^{-1} -
10^{-3}) j_0$. Here $j_0$ is in A/cm$^2$, and $v_m$ in cm/sec.
Note that the typical drift velocity is of order $10^{-4}j_0$
cm/sec.\cite{FBN}

The real part of the SW frequency, Eq. (\ref{okv}), also acquires
a term linear in the current, which is solely caused by the
anomalous Hall effect. The frequency of spin waves, with {\bf k}
non-parallel to the magnetization and to the current, is modified
by the current. The current increases or decreases the frequency,
depending on the direction of {\bf k}.

 When the wave-vector $\bf k$ and the current are along the
 magnetization,
 the critical velocity is given by
 \begin{equation}
 v_c(k)=\left[Dk + {\gamma (H_0+H_a)\over
 k}\right]\left(1+{\alpha\over2}\delta^2k^2\right). \label{vc0}
 \end{equation}
 $v_c(k)$ is minimal at $k=k_0$ given by
 \begin{equation}
 k_0^2
 ={1\over6\alpha\delta^2}[-(2+\alpha\beta)+\sqrt{4+28\alpha\beta
 +(\alpha\beta)^2}], \label{k0}
 \end{equation}
 where $\beta=\gamma (H_0+H_a)\delta^2/D$.

When the damping $\alpha$ is small, $\alpha\ll1/\beta$,
 the critical velocity coincides with the phase
velocity of the SW,\cite{GK} and $k_0$ is given by $k_0=[\gamma
(H_0+H_a)/D]^{1/2}$, $v_c(k_0)$ being equal to
\begin{equation}
v_c(k_0)=2\sqrt{D\gamma (H_0+H_a)}. \label{sk1}
\end{equation}
With typical values $D=(0.1 - 0.05)$ cm$^2$/sec and $\gamma
(H_0+H_a)=2\cdot10^9$ sec$^{-1}$,  one gets $v_c(k_0)\approx
3\cdot10^4$ cm/sec. Thus, the minimal critical current density is
$j_c=4\pi v_c(k_0)/ R_Mc\approx(3\cdot10^5 - 3\cdot10^7)
\mbox{A/cm}^2$.
 This
value of $j_c$ is by several orders  of magnitude smaller than
that obtained in Refs \onlinecite{FBN} and \onlinecite{LZ}.

The wave vector $k_0$ decreases with increase of $\alpha$.  If
$\alpha$ is large, $\alpha\gg1/\beta$, it follows from Eq.
(\ref{k0}), that $k_0$ does not depend on $D$ and $\gamma H_0$,
and is equal to: $k_0=(2/\alpha\delta^2)^{1/2}$. The critical
velocity in this case is
\begin{equation}
v_c(k_0)=\delta\gamma (H_0+H_a)\sqrt{2\alpha}, \label{sk2}
\end{equation}
increasing linearly with the external magnetic field.

 With $\sigma=10^5$ Om$^{-1}\cdot$
cm$^{-1}$, and $\omega_0= 10^{11}$ sec$^{-1}$, one gets
$\delta=10^{-4}$ cm. Then, with the above values of $D$ and
$\gamma (H_0+H_a)$ one finds that the inequality
$\alpha\gg1/\beta$ is fulfilled, if $\alpha$ is larger than
$10^{-2}$. The dependence of the minimal critical velocity on
$\alpha$ at different values of the conductivity is shown in Fig.
1.
\begin{figure}
\includegraphics[width=.90\columnwidth]{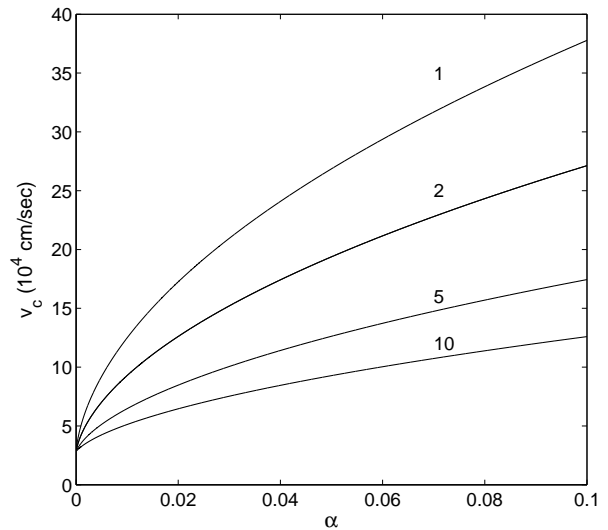}
 \caption{Dependence of the minimal critical velocity on $\alpha$
at different conductivities
$\sigma=(1,2,5,10)\times10^4(\Omega$cm)$^{-1}$, with parameters
$D=0.1$ cm$^2$/sec, $H_0+H_a=100$ Oe. Both {\bf k} and {\bf j}$_0$
are along the magnetization.}
\end{figure}

If {\bf k} is parallel to the current  and perpendicular to the
magnetization, Eqs. (\ref{vom}) and (\ref{nua}) yield
 \begin{equation}
 v_c(k)= {1\over k}\sqrt{\Omega_k(\Omega_k+\omega_0})\Bigl(1+{\alpha\delta^2k^2\over2}
 {2\Omega_k+\omega_0\over\Omega_k+\omega_0}\Bigr). \label{vcpi}
 \end{equation}
   Usually the inequality $\gamma
H_0\ll\omega_0$ holds. Then, at small $\alpha$ ,
$\alpha\ll1/\beta$, the minimal critical velocity is
 \begin{equation}
 v_c(k_0)=\sqrt{D\omega_0}\approx 10^5 \mbox{cm/sec}, \label{v01}
 \end{equation}
while the minimal critical current is of order $(10^6 - 10^8)$
A/cm$^2$. As before, the critical velocity increases with increase
of $\alpha$.
 When $\alpha$ is large, and satisfies the inequality $\alpha\gg1/\beta$,
 the minimal critical
 velocity is
 \begin{equation}
 v_c(k_0)=2\delta\sqrt{\gamma (H_0+H_a)\omega_0\alpha}, \label{vo2}
 \end{equation}
 The critical velocity and the critical current can be
 considerably smaller than the above values, if the ferromagnet is
 in a state
 close to an orientational phase transition caused  by an external
 magnetic field. We consider now the instability condition
 in several different arrangements of this type. In all cases we
 suppose that {\bf k} is along the dc current, since we are
 interested in the minimal critical current.

First, let in a uniaxial ferromagnet the external magnetic field
is aligned perpendicular to the easy axis, and $H_0$ is larger but
close to $H_a$. Then, the magnetization points along {\bf H}$_0$.
Repeating the previous calculation for this orientation of the
field, one gets for {\bf k} and {\bf j}$_0$ along the
magnetization:
\begin{equation}
v_c(k)={\omega_k\over k}\Bigl(1+{\alpha\delta^2k^2\over2}\Bigr),
\label{vcp}
\end{equation}
where \cite{ABP}
\begin{equation}
\omega_{\mathbf k}=\sqrt{[Dk^2+\gamma(H_0-H_a)](Dk^2+\gamma H_0)}.
\label{omp}
\end{equation}
At small $\alpha$
\begin{equation}
 \alpha\ll {D\over\gamma |H_0-H_a|\delta^2}, \label{al<1}
 \end{equation}
 and at
$H_0- H_a\ll H_a$, the critical velocity is minimum
 at a wave vector, given by
 \begin{equation}
 k_0=\Bigl[{\gamma^2H_a(H_0-H_a)\over D(D +\alpha\delta^2\gamma H_a)}\Bigr]^{1/4},
 \label{k0p}
 \end{equation}
and $v_c(k_0)$ is
\begin{equation}
v_c(k_0)=\sqrt{D\gamma H_a}, \label{vcop}
\end{equation}
which is considerably  smaller than the critical velocity
(\ref{sk1}).
 At large $\alpha$:
  $\alpha\gg D/\gamma(H_0-H_a)\delta^2$, the
 critical velocity
  equals to
 \begin{equation}
 v_c=\gamma\delta\sqrt{2\alpha H_a(H_0-H_a)}.
 \label{vop>}
 \end{equation}
Note that if $(H_0-H_a)/H_a$ is small, the inequalities Eq.
(\ref{al<1}) and $k_0^2\delta^2\ll1$ are fulfilled whenever
$\alpha$ is smaller than 1, i.e. the SW damping in this case
almost does not affect the critical current.

Suppose now that the ferromagnet is in a metastable state, with
the field ${\mathbf
 H}_0$  smaller than $H_a$ and opposite in direction to the
 magnetization.
 It follows than from Eq. (\ref{He}) that in Eq. (\ref{vc0}) and in all subsequent
 equations for $v_c$ the field
 $H_0$ should be replaced by $-H_0$. Hence, when $H_0$ approaches
 $ H_a$ the critical wave-vector and $v_c$ tend to zero. $k_0$ is
 restricted from below by the inequality $k_0^2\delta^2\gg1$. This
 gives for small damping $\alpha$, which satisfies the inequality (\ref{al<1}):
  $\gamma(H_a-H_0)\gg
 D\delta^{-2}$,  $v_c\gg2D/\delta\approx 10^3$ cm/sec,
 and $j_c\gg 10^4$ A/cm$^2$. This implies that a relatively small
 current of order or larger than $(3-5)\times 10^4$ A/cm$^2$ can drive the
 magnetization switching at magnetic fields smaller than $H_a$,
 if the current flows along the magnetization.
  Note that, as in the previous case,  the inequality Eq. (\ref{al<1}) is
  equivalent to the condition $\alpha\ll1$.

Finally, consider thin films, with $kd\ll1$, $d$ is the film
width.  The SW spectrum in this case at different directions of
the magnetization and external magnetic field was derived in many
papers, see e.g. Refs. \onlinecite{EM}.  We consider the case,
when the external magnetic field  is perpendicular to the film
plane and larger than $4\pi M_0 +H_s -H_a$, where $H_a$ is the
volume anisotropy field, the easy axis being in a symmetry
direction of the film, and $H_s$ is the surface anisotropy field.
Then, the magnetization is also perpendicular to the film, and the
spectrum of SW with {\bf k} in the plane is:
\begin{equation}
\omega_k=\sqrt{[\gamma(\tilde{H} +2\pi M_0kd)
+Dk^2][\gamma\tilde{H} +Dk^2]}, \label{omf}
\end{equation}
where $\tilde{H}=H_0+H_a-H_s-4\pi M_0$.

As argued above, the damping can be neglected if the ferromagnet
is in the vicinity of the phase transition. Then the critical
velocity is equal to the SW phase velocity, and is minimal at
$k_0=\sqrt{\gamma\tilde{H}/ D}$, while $\tilde{H}$ should be
larger than $D/\gamma\delta^2$. The minimal critical velocity is
given by:
\begin{equation}
v_c(k_0)=\left(4\gamma\tilde{H}D+
\omega_0d\sqrt{\gamma\tilde{H}D}\right)^{1/2}. \label{v02}
\end{equation}
The above inequalities yield that $v_c(k_0)$ is restricted from
below as: $v_c>\sqrt{\omega_0Dd/\delta}\approx 10^4$ cm/sec.
Hence, the critical current is larger than $(10^5 - 10^7)$
A/cm$^2$.

In conclusion, we have calculated the effect of an electric
current on the SW spectrum in a ferromagnetic metal. We have shown
that the ordinary and anomalous Hall currents lead to the
reduction of the SW damping, caused by the eddy currents. At
sufficiently strong currents the damping changes sign, and a SW
instability develops. The critical current of the instability is
determined mainly by the anomalous Hall effect, and may be much
smaller than the critical current for SW excitation with
spin-transfer torques.

The author thanks A. Gerber for helpful discussions. This research
was supported by the Israel Science Foundation grant No. 633/06.


\begin{thebibliography}{99}
\bibitem{TBB} Y. Tserkovnyak, A. Brataas, G.~E.~W. Bauer, and
B.~I. Halperin, Rev. Mod. Phys. {\bf 77}, 1375 (2005).
\bibitem{Sl1} J.~C. Slonczewski, J. Magn. Magn. Mater {\bf 159},
L1 (1996); J.~C. Slonczewski, J. Magn. Magn. Mater {\bf 195}, L261
(1999).
\bibitem{Ber1} L. Berger, Phys. Rev. B {\bf 54}, 9353 (1996)
\bibitem{BJZ} Ya.~B. Bazaliy, B.~A. Jones, and S.-C. Zhang,
Phys. Rev. B {\bf 57}, R3213 (1998).
\bibitem{FBN} J. Fern\'{a}ndez-Rossier, M. Braun, A.~S.
N\'{u}\~{n}ez, and A.~H.~MacDonald, Phys. Rev. B {\bf 69}, 174412
(2004).
\bibitem{LZ} Z. Li and S.Zhang, Phys. Rev. Lett. {\bf 92}, 207203
(2004).
\bibitem{TSB} Y. Tserkovnyak, H.~J. Skadsem, A. Brataas, and
G.~E.~W. Bauer, Phys. Rev. B {\bf 74}, 144405 (2006).
\bibitem{PB} M.~L. Polianski and P.~W. Brouwer, Phys. Rev. Lett.
{\bf 92}, 026602 (2004).
\bibitem{JCS} Y. Ji, C.~L. Chien, and M.~D. Stiles, Phys. Rev. Lett.
{\bf 90}, 106601 (2003). M.~D. Stiles, J. Xiao and A. Zangwill,
Phys. Rev. B {\bf 69}, 054408 (2004).
\bibitem{XYO} H. Xi, Y. Yang, J. Ouyang, Y. Shi, and K.-Z. Gao,
Phys. Rev. B {\bf 75}, 174411 (2007).
\bibitem{ABP} A.~I. Akhiezer, V.~G. Bar'yakhtar, and S.~V.
Peletminskii, {\it Spin Waves} (North-Holland, Amsterdam, 1969).
\bibitem{GK} L.~\'{E} Gurevich and I.~Ya. Korenblit, Zh. Eksp.
Teor. Fiz. {\bf 48}, 652 (1965) [Sov. Phys. JETP {\bf 21}, 431
(1965)].
\bibitem{MAF} T. Miyasato, N. Abe, T. Fujii, A. Asamitsu, S.
Onoda, Y. Onose, N. Nagaosa, and Y. Tokura, Phys. Rev. Lett. {\bf
99}, 086602 (2007).
\bibitem{Gerber} A. Gerber, A. Milner, A. Finkler, M. Karpovski, L.
Goldsmith, J. Tuaillon-Combes, O. Boisron, P. M\'{e}linon, and A.
Perez, Phys. Rev. B {\bf 69}, 224403 (2004).
\bibitem{EM} R. Arias and D.~L. Mills,
Phys. Rev. B {\bf 60}, 7395 (1999); S.~M. Rezende, F.~M. de
Aguiar, and A. Azevedo, Phys. Rev. B {\bf 73}, 094402 (2006).
\end{thebibliography}
\end{document}